# Kinetics of the lattice response to hydrogen absorption in thin Pd and CoPd films.

S. S. Das, G. Kopnov and A. Gerber

Raymond and Beverly Sackler Faculty of Exact Sciences,

School of Physics and Astronomy, Tel Aviv University,

Ramat Aviv, 69978 Tel Aviv, Israel

Hydrogen can penetrate reversibly a number of metals, occupy the interstitial sites and cause large expansion of the crystal lattice. The question discussed here is whether the kinetics of the structural response matches hydrogen absorption. We show that thin Pd and CoPd films exposed to a relatively rich hydrogen atmosphere (4% $H_2$) inflate irreversibly, demonstrate the controllable shape memory, and duration of the process can be orders of magnitude longer than hydrogen absorption. The dynamics of the out-of-equilibrium plastic creep is well described by the Avrami - type model of the nucleation and lateral domain wall expansion of the swelled sites.

Corresponding author: A. Gerber, email: gerber@tauex.tau.ac.il

# Introduction.

An anticipated transition to hydrogen as the main ecologically clean mobile energy source [1, 2] requires profound knowledge and understanding of the hydrogen – solid matter interactions for the treatment, storage and monitoring tasks. Hydrogen can penetrate reversibly a number of metals (Pd, Nb, Ti, Mg, V, etc.), occupy the interstitial sites and cause a large expansion of the crystal lattice [3]. The question that remains surprisingly open after decades of research is whether the kinetics of the hydrogen absorption and that of the structural response match each other. Coupling between the hydrogen migration and the lattice expansion is an important element in understanding the processes of hydrogenation and of the theoretical modelling of hydrogen diffusivity [4, 5]. Despite an obvious importance of this assumption, there is little evidence on the time-dependent correlation between the two processes, mainly because of an experimental challenge to monitor each of them independently. Hydrogen atoms are invisible by the majority of lattice characterization tools, excluding the neutron scattering [6, 7]. On the other hand, the techniques used to study hydrogen diffusion [8] do not provide independent information on the structural evolution. In the absence of verification, the structural and strain time-dependent data are frequently taken as the information source on hydrogen diffusion, hydride phase transformations and their spatial distribution [9 - 11]. The assumption of the kinetics coupling is not obvious. Multiple out-of-equilibrium processes, where relaxation from a metastable state lags behind the state buildup, are known. Here, we demonstrate that such an out-of equilibrium process can occur in hydrogenated materials and that the kinetics of the structural response to the stress generated by hydrogen accommodation can differ significantly from that of hydrogen diffusion.

For the reasons clarified in the following, we studied two high-resistivity hydrogen-absorbing systems: thin films of palladium, where resistivity is enhanced by surface scattering and fractal topology, and CoPd alloys, where resistivity increases with Co concentration. Ferromagnetic CoPd alloys and multilayers attracted attention recently as materials for high-sensitivity magnetic detection of hydrogen using ferromagnetic resonance [12] and the extraordinary Hall effect [13, 14].

**Experimental**

Experimental disentanglement between hydride formation and the lattice response has been achieved by a simple electric resistance measurement. Absorption of hydrogen increases the resistivity of bulk Pd [15 - 18] and Pd-rich alloys [19] by a value $\Delta\rho_H$ that depends on the concentration of the absorbed hydrogen and the composition and structure of the material. $\Delta\rho_H$ is of few μΩcm in the Pd β-hydride state at room temperature [20]. Another contribution to the resistance change, overlooked in previous studies, is due to the expansion of thin films thickness. Thin films grown on rigid substrates cannot expand laterally within the film plane due to adhesion to the surface. Suppression of the in-plane expansion is equivalent to application of the in-plane compressive stress leading to the out-of-plane expansion enhanced by Poisson's effect. The out-of-plane elastic expansion of the [111] textured Pd film is about 12.6% when the atomic ratio between Pd and the absorbed hydrogen is 1 [21]. The change in resistivity between the hydrogen-free state with resistance:

$$R_0 = \frac{\rho_0}{T_0} \cdot \frac{l}{w}$$

and the hydrogenated state with:

$$R_1 = \frac{\rho_1}{T_1} \cdot \frac{l}{w}$$

is:

$$\Delta\rho = \frac{1}{1+\gamma}(\Delta\rho_H - \gamma\rho_0) \qquad (1)$$

where: $\rho_0$ is the initial resistivity, $T_0$ is the nominal thickness, and $l$ and $w$ are the length and width respectively, of the film that don't change during the hydrogenation because of adhesion to the substrate. Thickness of the hydrogenated film is: $T_1 = T_0 + \Delta T = (1+\gamma)T_0$, where $\gamma = \Delta T/T_0$ is the thickness expansion coefficient (strain). For simplicity, we assume that resistivity follows Matthiessen's rule: $\rho_1 = \rho_0 + \Delta\rho_H$, and the initial $\rho_0$ is not affected by strain.

The geometrical (thickness expansion) term $\gamma\rho_0$ depends on the initial resistivity of the material. For thick Pd films with a resistivity of 10 - 15 μΩcm, depending on the microstructure and the hydride term $\Delta\rho_H \approx 6$ μΩcm [20], the negative geometrical contribution is relatively small. However, in high resistivity films, where $\Delta\rho_H \ll \gamma\rho_0$, the geometrical term is dominant, the overall resistance response to hydrogen loading is negative, and $\Delta\rho \propto -\rho_0$. Reversal of the resistance response polarity from positive to negative at the critical resistivity threshold of about 50 μΩcm in 4% $H_2$ atmosphere has been demonstrated in thin Pd, thick $PdSiO_2$ granular mixtures and CoPd alloy films with variable resistivity [22]. Earlier observations of the reduced resistance in the hydrogenated state were attributed to the lateral swelling of disconnected Pd clusters [23 – 27], contribution of hydrogen electrons to the conducting band [28], and even to the onset of room temperature superconductivity [29]. It seems, however, that inflation of a film thickness provides a simple and consistent explanation of the observed effects in both continuous and discontinuous hydrogenated films [22]. We adapt this interpretation and, in the following, distinguish between hydride formation and lattice expansion by the respectively positive and negative terms in the resistance response to hydrogen loading. The distinction among them is clear when the reversibility and the time dependence of the two processes are different.

The samples used in this study were 3nm to 15 nm thick polycrystalline Pd and $Co_xPd_{100-x}$ alloy films with lateral dimensions 5 x 5 mm grown by rf-magnetron sputtering onto room-temperature glass substrates. Binary $Co_xPd_{100-x}$ films with Co atomic concentrations in the range $0 \leq x \leq 80$ were co-sputtered from separate targets (1.3" diameter and 2mm thick). Co and Pd are soluble and form an equilibrium fcc solid solution phase at all compositions during the room temperature deposition. No post-deposition annealing was done. The desired composition and thickness were controlled by the relative sputtering rates in the range 0.01 – 0.1 nm/sec with the respective rf power between 0 and 85 W and tested by EDS (energy-dispersive X-ray spectroscopy) measurements. Resistance was measured using the Van der Pauw protocol. Electrical contacts were attached by bonding Al/Si wires. The setup was equipped with a gas-control chamber, which enabled performing measurements at variable hydrogen concentrations. The hydrogen-induced resistance changes were extracted from measurements performed in dry nitrogen and in 4% $H_2/N_2$ mixture gas at 1 atm pressure at room temperature. The 15 nm thick and thinner Pd films were below the delamination thickness threshold [30] and were stable under repeated hydrogenation cycles. No buckling was observed in the CoPd samples at all tested thicknesses. The 15 nm thick films formed a continuous metallic layer. The 3 nm films had inhomogeneous meandric morphology with continuous metallic paths across the sample [22].

**Results and Discussion.**

Figure 1 presents the resistivity response to a sequence of hydrogenation and dehydrogenation cycles (sequential exposure to 1 atm 4% $H_2/N_2$ gaseous mixture followed by $N_2$) of four 15 nm thick samples: pure Pd (a) and three $Co_xPd_{100-x}$ alloys with x=15 (b), x=30 (c) and x=50 (d). The starting resistivity in $N_2$ is the lowest in Pd film (22 $\mu\Omega$cm) and grows gradually in alloy samples with increasing Co content to 108 $\mu\Omega$cm in $Co_{50}Pd_{50}$. The first exposure to hydrogen varies qualitatively with increasing Co content and initial resistivity, resistance increases sharply and saturates in the low resistivity Pd (a), increases sharply and decreases in (b), and decreases in samples (c) and (d) with higher resistivity ($\rho_0 > 50$ $\mu\Omega$cm). Removal of hydrogen is similar in all samples: resistance drops and saturates in $N_2$. Starting from the second – third cycle the resistance response becomes reproducible and similar in all samples: resistance increases sharply when exposed to hydrogen and drops when hydrogen is removed (it will become clear in the following that reproducibility is achieved after a long enough exposure to hydrogen and not due to the number of cycles). The final resistance in $N_2$ is lower than the starting one in all samples. The entire sequence is a composition of reproducible rapid increase/decrease responses to the loading/unloading of hydrogen superposed with an irreversible gradual reduction in resistivity. Following the model above, the sequence can be interpreted as a superposition of reversible hydride formation-removal signals on the background of the irreversible thickness inflation, while the relative magnitude and time duration of the latter increases with Co content. The overall resistance response of low resistivity Pd film (a) is dominated by the hydrogenation scattering contribution, while that of the high resistivity $Co_{50}Pd_{50}$ sample (d) by the irreversible thickness expansion.

The difference in the kinetics of hydride formation and thickness inflation in pure Pd is demonstrated in Fig.2 for two identical 3 nm thick samples with a resistivity of 750 µΩcm (this high resistivity is due to the vicinity to the conductance percolation threshold). The first sample was exposed continuously to hydrogen (solid line) while the second to a sequence of hydrogen loadings and removals (open circles). The immediate resistance increase on the hydrogen filling and drop on the respective removal (sample 2), interpreted as the hydride scattering term, indicate a rapid hydrogen diffusion into and out of the metal. The sequence of the hydride terms is superposed with a slowly decreasing background. The irreversible reduction in resistance is equal in both samples, therefore the final thickness expansions are identical. Two points are notable: 1) the lattice response is much slower than the hydrogen diffusion in and out of the material; 2) the lattice expansion is frozen when hydrogen is extracted (resistance is constant in $N_2$), and the process of expansion is recovered from the same state when the hydride is recovered (resistance increases on reloading to the same value prior to hydrogen extraction. See Fig.2 inset). Thus, the system demonstrates the shape memory. It is worth noting that partial thickness recovery is observed when films are flushed in air instead of nitrogen.

The time dependence of the normalized resistance changes due to hydride formation $\frac{\Delta R_H}{\Delta R_{H,max}}$ and the thickness expansion $\frac{\Delta R_T}{\Delta R_{T,max}}$ are shown in Fig.3a and 3b respectively for a number of 15 nm thick CoPd films with different Co concentrations, including a pure Pd film. $\Delta R_{H(T),max}$ are the largest saturated values of the respective changes for each sample. Fig. 3c presents the effective time of each process $t_{H,50}$ and $t_{T,50}$ defined as the time at which the respective resistance term changed by half. Hydride formation is accomplished within a few tens to hundred seconds in all samples. In alloys, the process of hydride formation is quicker with increasing Co content: $t_{50}$ = 60 sec in $Co_{10}Pd_{90}$ down to a few sec in $Co_{40}Pd_{60}$. The increase in the diffusion rate in diluted Pd alloys is consistent with that reported for NiPd alloys [31]. The thickness expansion time scale is entirely different: from 10 secs in the x=20 sample to $10^5$ sec in the x=80 one. $t_{T,50}$ can be approximated by: $t_{T,50} \propto e^x$ (solid line in Fig.3c), i.e. it increases exponentially with the concentration of Co.

The shape of the time dependent $\Delta R_T(t)$ (Fig.3b) is informative. At low Co content, $\Delta R_T$ drops immediately with exposure to hydrogen. The creep rate $dR_T/dt$ is intuitively clear: it is at its maximum at the beginning when stress is the highest, and decreases with time when stress is released gradually by plastic deformations. Co-rich samples demonstrate a different behavior: an onset of expansion occurs after a long delay and the relaxation curve has a characteristic S-shape (see Fig.1d in a linear time scale). Such dynamics can be understood in the framework of the Avrami or the Johnson-Mehl-Avrami-Kolmogorov (JMAK) model [32 – 37], which was first formulated to describe kinetics of isothermal recrystallization of metals. The transformation proceeds by nucleation and growth of a new phase, and can be summed up by a simple formula: $V(t) = 1 - e^{-V_e(t)}$, where V is the fraction of the transformed phase and $V_e$ is the so-called extended volume of the transformed phase, that is the volume the transformed phase would acquire if the overlap among the growing nuclei was disregarded. In general, the model can describe any non-coherent transition from a metastable state to the lowest energy equilibrium state by a sequence of local transition events occurring when energy barriers prevent an immediate global transition to the equilibrium. The phenomenology appeared to be quite universal and the model was used, among others, in describing the kinetics of thin film growth [38], phase transition in ferroelectrics [39, 40], magnetization reversal in ferromagnets [41 – 43], distribution of infections in networks [44] and evolution of religions [45]. In ferromagnets it was used to describe the reversal of magnetization in the magnetized films subjected to an external magnetic field antiparallel to the magnetization vector. The equilibrium state is the one in which the magnetization is oriented parallel to the applied field, and the metastable one is when the magnetization and field are antiparallel. If the field value is smaller than required to overcome the magnetic anisotropy energy barrier, the magnetization reverses not by a coherent rotation in the entire volume but rather by a sequence of distributed non-coherent local events. The Fatuzzo-Labrune theory [39, 41] describes the process by two microscopic phenomena: nucleation of new domains with reversed magnetization and their expansion by domain wall propagation. In the case of the hydrogenated films, the metastable state can be created when hydrogen atoms diffuse into the material, occupy the interstitial states and generate an internal pressure on the lattice, while an immediate expansion is prevented by either the thermodynamic constrains, such as the phase transition between the α and β states, or/and by the structural ones, such as adhesion to a substrate. In analogy with the Fatuzzo-Labrune theory [39, 41], we model the process of thickness deformation by nucleation of new "swelled" domains and their lateral expansion due to domain wall propagation. These are described by the probability of nucleation per unit time $p_n$ and by the effective domain wall velocity $v$. The entire process can be characterized by a dimensionless parameter $k$ defined as:

$$k = \frac{v}{p_n r_n}$$

in which $r_n$ is the radius of a nucleation site. Temporal variation of resistance is analytically simple in two limiting cases: one in which the nucleation rate dominates and $k \ll 1$ and another in which the domain wall propagation is dominant and $k \gg 1$:

$$\frac{\Delta R_T}{\Delta R_{T,max}} = \begin{cases} \exp\left[-\frac{k^{-2}(p_n t)^3}{3}\right] - 1 & k \gg 1 \\ \exp(-p_n t) - 1 & k \ll 1 \end{cases} \qquad (2)$$

The Fatuzzo-Labrune model does not account for the final stage of the process when the last stressed regions are annihilated. This final stage can be fitted by an additional exponential term $\exp(-p_a t)$ where $p_a$ is the site annihilation probability [42, 43]. Dimensions of the nucleation and annihilation sites are assumed to be similar. The entire creep process in terms of the varying effective thickness $\Delta T(t)$ is given by:

$$\frac{\Delta T(t)}{\Delta T_{max}} = 1 - \frac{\Delta R_T(t)}{\Delta R_{T,max}} = \begin{cases} 1 - \left\{A\exp\left[-\frac{k^{-2}(t/\tau_n)^3}{3}\right] + B\exp(-t/\tau_a)\right\} & k \gg 1 \\ 1 - [A\exp(-t/\tau_n) + B\exp(-t/\tau_a)] & k \ll 1 \end{cases} \qquad (3)$$

$$A + B = 1$$

where $\tau_n = 1/p_n$ and $\tau_a = 1/p_a$ are the effective relaxation and annihilation time respectively.

Fig.4 presents a convincing fitting of Eq.3 (solid lines) to the time dependent data (symbols) for the 3 nm thick Pd film (a) and two 15 nm alloy samples: $Co_{35}Pd_{65}$ (b) and $Co_{70}Pd_{30}$ (c). The first two are well fitted in the $k \ll 1$ limit (Pd: $\tau_n=5\times10^2$ sec and $\tau_a=8\times10^3$ sec; $Co_{35}Pd_{65}$: $\tau_n=1\times10^2$ sec and $\tau_a=1\times10^3$ sec), while $Co_{70}Pd_{30}$ fits the $k \gg 1$ one ($k=95$, $\tau_n=6\times10^4$ sec and $\tau_a=1\times10^5$ sec), in which the nucleation is rare and the incubation time is long compared with lateral expansion of the nucleated domains.

The conclusions of this work are based on a given interpretation of the resistance data. Credibility of the interpretation is supported by the results obtained by other techniques reported in the literature. Irreversible changes in the crystalline structure and evidence of plastic out-of-plane expansion in thin Pd films during the β-phase formation were found by XRD [21, 46, 47] and microcantilever [48] measurements. Similar irreversible expansion of thickness was also observed in other hydrogenated metals as well: Nb films [49], Mg-Y, Mg-Ni [50] and PdAg [51] alloys. Exponential time dependent thickness swelling was found in Nb films exposed to hydrogen by STM measurements, while the time scale exceeded by far the values expected for hydrogen diffusion time [49]. The divergence of the plastic deformation time in Co-rich samples is consistent with the superior creep stability of fcc Co-based superalloys [52].

To summarize, the kinetics of hydrogen penetration into metals can differ significantly from the respective lattice response. Hydrogenation of Pd and CoPd films in 4% $H_2/N_2$ atmosphere creates an out-of-equilibrium state in which stress is built up rapidly with hydrogen absorption and is released by a slow plastic thickness growth. Expansion of thickness is irreversible. The creep can be frozen by removal of hydrogen and restarted from the frozen state by hydrogen reloading. Thus, the material demonstrates the shape memory. The dynamics of the creep can be described by the Avrami-type model with the nucleation of isolated swelled sites and lateral domain wall expansion. The nucleation rate slows down in Co-rich alloys, such that an effective thickness inflation can occur 30 hours after the exposure to hydrogen, five orders of magnitude slower than the gas diffusion time.

## Acknowledgements.

The work was supported by the Israel Science Foundation grant No. 992/17.

**Figure captions.**

Fig.1. Resistivity response to a sequence of hydrogenation and dehydrogenation cycles (sequential exposure to a 1 atm 4% $H_2/N_2$ gaseous mixture followed by $N_2$) of four 15 nm thick samples: pure Pd (a) and three $Co_xPd_{100-x}$ alloys with x=15 (b), x=30 (c) and x=50 (d). The initial atmosphere is $N_2$.

Fig.2. Time dependent normalized resistance of 3 nm thick Pd exposed continuously to 4% $H_2/N_2$ mixture (solid red line) and to a sequence of hydrogenation and dehydrogenation cycles (open circles). Inset: zoom on one of the $H_2$-$N_2$ cycles. Resistance is frozen in $N_2$ and decreases in $H_2$.

Fig.3. Time dependence of the normalized resistance changes due to hydride formation $\frac{\Delta R_H}{\Delta R_{H,max}}$ (a) and thickness expansion $\frac{\Delta R_T}{\Delta R_{T,max}}$ (b) of 15 nm thick $Co_xPd_{100-x}$ alloys with different Co content x. (c) The half-time of hydride formation $t_{H,50}$ (left vertical axis) and of thickness expansion $t_{T,50}$ (right vertical axis) as a function of Co concentration x. Lines are guides for the eye.

Fig.4. Normalized change of thickness as a function of the hydrogen exposure time of 3 nm thick Pd (a), 15 nm thick $Co_{35}Pd_{65}$ (b) and 15 nm thick $Co_{70}Pd_{30}$ (c) films. Symbols represent the experimental data and lines are fitting by Eq.3. Samples (a) and (b) were fitted in the $k \ll 1$ limit and sample (c) in the $k \gg 1$ one.

**Figures**

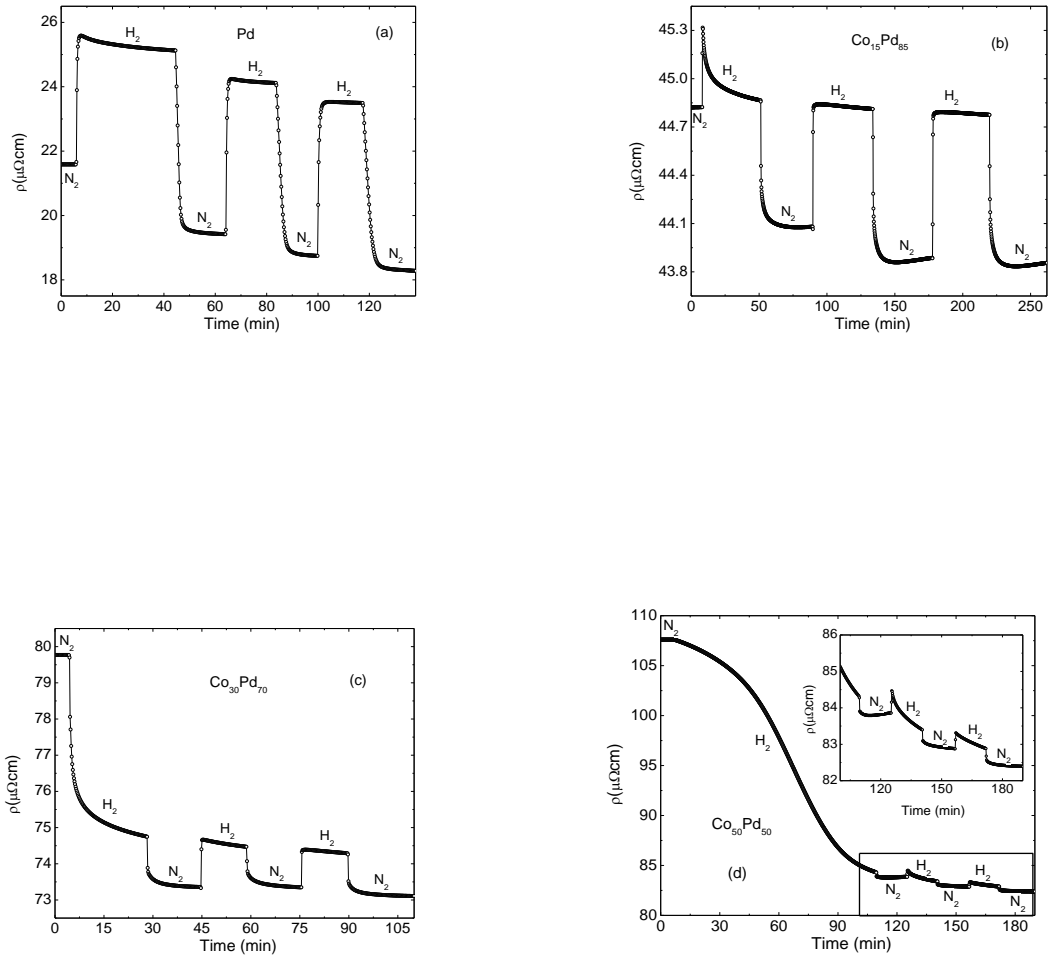

Fig. 1

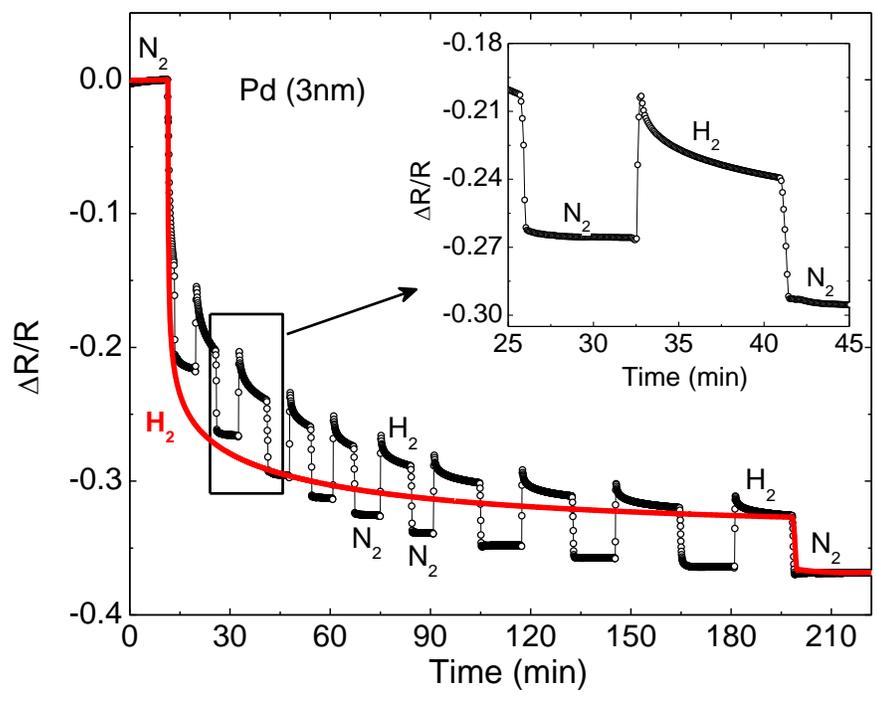

Fig. 2

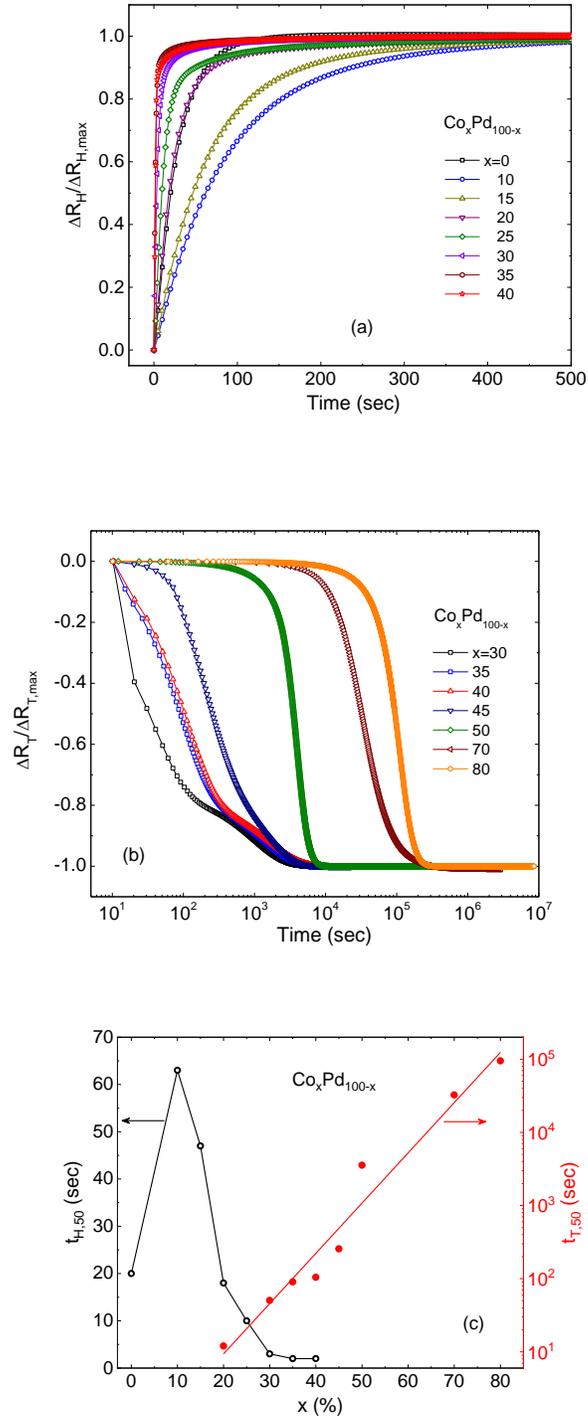

Fig. 3

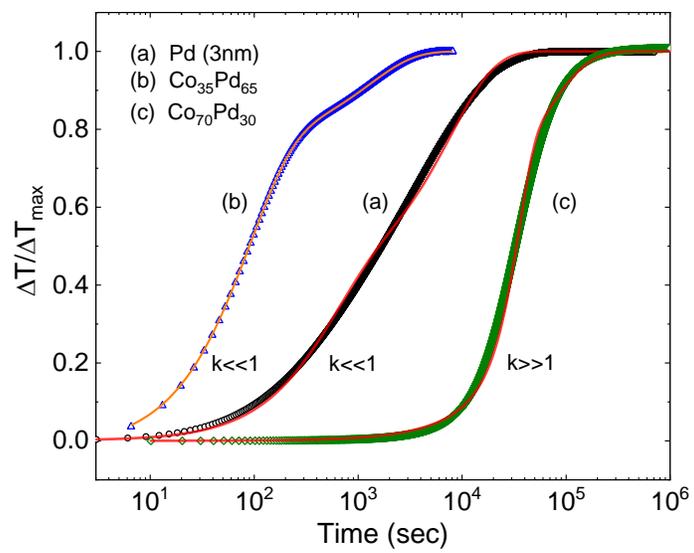

Fig. 4